\documentclass[11pt]{article}
\usepackage{moriond,epsfig}

\newcommand{\gev}{\,\mbox{GeV}}
\newcommand{\ov}{\overline}
\newcommand{\lt}{\left}
\newcommand{\rt}{\right}

\newcommand{\eq}[1]{(\ref{#1})}

\newcommand{\mg}[1][]{\ensuremath{M_{\tilde{g}}^{#1}}}
\newcommand{\msb}[2][]{\ensuremath{m_{\tilde{b}_{{#2}}}^{{#1}}}}
\newcommand{\tb}[1][]{\ensuremath{\tan^{#1}\!\beta}}

\begin{document}
\vspace*{2cm}
\boldmath
\title{\mbox{\normalsize CERN-TH/2001-210} \hfill 
       \mbox{\normalsize hep-ph/0107333}\\[3mm]
Supersymmetric corrections to Higgs decays and $b\to s \gamma$
  for large $\tan \beta$~\footnote{Talk at the \emph{Rencontres de
    Moriond: electroweak interactions and unified theories}, 10-17 Mar
  2001, Les Arcs 1800, France.}} 
  
\unboldmath

\author{Ulrich Nierste}

\address{CERN -- TH, 1211 Geneva 23, Switzerland}

\maketitle\abstracts{
If $\tan \beta$ is large, supersymmetric QCD corrections can 
become large, putting naive perturbation theory into doubt. I show how
these $\tan \beta$-enhanced corrections can be controlled to all
orders in $\alpha_s \tan\beta$. The result is shown for the decays 
$H^+ \to t \ov{b}$ and $b \to s \gamma$.
}

\section{Large corrections to all orders}
The Minimal Supersymmetric Standard Model (MSSM) contains two Higgs
doublets $H_u$ and $H_d$. Their neutral components acquire the vacuum
expectation values $v_u$ and $v_d$ with $v\equiv
\sqrt{v_u^2+v_d^2}=174\; \gev $. Recently scenarios with large $\tan
\beta \equiv v_u/v_d$ (corresponding to $v_d \ll v_u \simeq 174\; \gev
$) have attracted increasing attention: this region of the parameter
space is experimentally least constrained by the bounds from Higgs
searches at LEP~\cite{lep}. A theoretical motivation for large $\tan
\beta$ scenarios stems from GUT theories with bottom--top Yukawa
unification, which require $\tan \beta = {\cal O} (50)$
\cite{gut,gut2}. At tree-level right-handed down-quark fields do not
couple to $H_u$ and the bottom quark mass is related to the
corresponding Yukawa coupling $h_b$ by $m_b = h_b v_d = h_b v \cos
\beta$. For large $\tan \beta$ this has two important consequences:
first $h_b$ is large, of order 1. Second, radiative corrections to the
couplings of Higgs bosons to $b$ quarks proportional to $h_b \sin
\beta$ can occur. These are enhanced by a factor of $\tan \beta$
compared to the tree-level result and stem from the supersymmetry
breaking terms. The dominant $\tan \beta$-enhanced corrections are
supersymmetric QCD (SQCD) contributions, i.e.\ loop diagrams with
squarks and gluinos.

Observables can be affected by these large corrections in two ways:
\begin{itemize}
\item[I] The leading order contribution is proportional to $\cot
  \beta$. This suppression is lifted in the loop corrections of the
  next-to-leading order (NLO).  
\item[II] $\tan \beta$-enhanced corrections enter the counterterms,
  which appear in higher order corrections.
\end{itemize}
An example for type-I corrections is the $H^+ \ov{t}_R s_L$ coupling
$\propto m_t \cot \beta $ appearing in the one-loop matrix element for
$b \to s \gamma$.  SQCD vertex corrections are not suppressed by
$\cot\beta$, so that the two-loop matrix element is $\tan
\beta$-enhanced. Clearly, three- or more loops cannot produce more
factors of $\tan \beta$, because the bare lagrangian contains $h_b$
and $\beta$ only in the combinations $h_b \cos \beta$ and $h_b \sin
\beta$. The enhancement mechanism of type-II is related to the
renormalization of $h_b$. SQCD corrections induce a counterterm to
$h_b^{\mathrm{tree}}=m_b/(v \cos \beta)$ which reads $\delta
h_b=\delta m_b/(v \cos \beta)$ in terms of the mass counterterm
$\delta m_b$. Also $\delta m_b$ contains terms proportional to $h_b v
\sin \beta$, so that $\delta h_b$ is $\tan \beta$-enhanced. Writing
$\delta m_b=-m_b \Delta m_b$ one finds
\begin{equation}
h_b = h_b^{\mathrm{ren}} + \delta h_b 
    = h_b^{\mathrm{tree}}   \lt( 
      1 - \Delta m_b \rt)  \label{h1}
\end{equation}
at the one-loop level. In this talk I consider the on-mass-shell
renormalization scheme, in which $\delta m_b$ is adjusted to cancel
the on-shell self-energy of the $b$ quark. For other schemes I refer
to \cite{cgnw}. Then $\Delta m_b$ reads \cite{gut2}:
\begin{eqnarray}
  \Delta m_b &=& \frac{2\alpha_s}{3\pi}\mg \,
                        \mu\, \tb\, I(\msb{1},\msb{2},\mg) 
   \; \approx \;  \frac{\alpha_s}{3\pi}
                  \frac{\mu \mg}{M_{\tilde{g},\tilde{b}_{1,2}}^2}\, \tb \, .
   \label{dmb}
\end{eqnarray}
Here $\msb{1,2}$ and \mg\ are the $b$ squark and gluino masses and
$\mu$ is the Higgsino mass parameter. The last formula in \eq{dmb} is
approximate with $M_{\tilde{g},\tilde{b}_{1,2}}$ being the average of
$\msb{1}$,$\msb{2}$ and \mg.  The exact formula contains \cite{gut2}
\begin{equation}
  I(a,b,c) = \frac{1}{(a^2-b^2)(b^2-c^2)(a^2-c^2)}
  \left(a^2b^2\log{\frac{a^2}{b^2}}
        +b^2c^2\log{\frac{b^2}{c^2}}
        +c^2a^2\log{\frac{c^2}{a^2}}\right)\,.
\end{equation}
By inspecting the one-loop SQCD corrections to Higgs and top decays
calculated in \cite{SQCDHtb} one verifies that their $\tan
\beta$-enhanced portion indeed stems from the counterterm diagram
involving $\delta h_b$. The next-to-leading order result for $b\to
s \gamma$ \cite{cdgg} in addition contains type-I corrections 
from the loop-corrected $H^+ \ov{t}_R s_L$ vertex. 

In \cite{gut2} the type-II corrections have been derived in a
different way, by investigating the loop-induced coupling of 
$b$ quarks to the other Higgs doublet $H_u$. This amounts to replacing
$h_b$ by  
\begin{equation}
h_b^\mathrm{eff} 
    = \frac{ h_b^{\mathrm{tree}} }{  
      1 + \Delta m_b } . \label{h2}
\end{equation}
The scope of all analyses \cite{gut2,SQCDHtb,cdgg} were one-loop
corrections to $h_b$, and obviously \eq{h1} and \eq{h2} agree up to
terms of order $(\Delta m_b)^2 = {\cal O} (\alpha_s^2) $.  The two
methods are related by a Ward identity. Since $\Delta m_b$ can be of
order 1, there can be drastical numerical differences between \eq{h1}
and \eq{h2}.  At first glance this raises doubts whether perturbation
theory works if $\tan \beta$ is large.

In \cite{cgnw,cgnw2} it has been investigated how higher orders are
affected by these enhanced SQCD corrections: unlike type-I corrections
the type-II corrections from $\delta h_b$ in \eq{h1} appear
recursively in \emph{all}\ orders of perturbation theory. $\delta h_b$
also enters the renormalized $\tilde b_L$--$\tilde b_R$ mixing in the $b$
squark mass matrix. As a consequence enhanced contributions to $\delta
m_b$ at order $\alpha_s^n$ appear from one-loop self-energy diagrams
in which a $\tilde b_L$--$\tilde b_R$ flip stems from the ${\cal O}
(\alpha_s^{n-1})$ term of $\delta h_b$ \cite{cgnw}.  The enhanced
higher order corrections simply sum to a geometric series \cite{cgnw}:
\begin{equation}
h_b = h_b^{\mathrm{ren}} + \delta h_b 
    = h_b^{\mathrm{tree}} \lt( 
      1 - \Delta m_b + (\Delta m_b)^2 - (\Delta m_b)^3 + \ldots  \rt)
    = \frac{h_b^{\mathrm{tree}}}{1+\Delta m_b} . \label{heff}
\end{equation}
The procedure of \cite{gut2} directly arrives at this result, as
evidenced by \eq{h2}. Indeed, $\tan \beta$-enhanced corrections are
absent in the higher order contributions to the loop-induced $\ov{b}b
H_u$ vertex. The proof of this feature and the establishment of the
geometric series in \eq{heff} involves power counting, the operator
product expansion and standard infrared theorems \cite{cgnw}. 

\boldmath
\section{$H^+ \to t \ov{b}$ and $b \to s \gamma$}
\unboldmath Next I present the effect of the resummation in \eq{heff}
on two decay rates. The first plot shows the relative size of the SQCD
corrections in the decay rate $\Gamma (H^+ \to t \ov{b}) $ for two
sets of parameters: $M_{H^+}=350\,$GeV, $M_{\tilde{g}}=500\,$GeV,
$M_{\tilde{b}_2}=200\,$GeV ($<M_{\tilde{b}_1}$),
$M_{\tilde{t}_2}=180\,$GeV ($<M_{\tilde{t}_1}$)(``light'') and
$M_{\tilde{g}}=M_{\tilde{b}_2}=M_{\tilde{t}_2}=1000\,$GeV (``heavy'')
\cite{cgnw}:

\begin{center}  
\psfig{figure=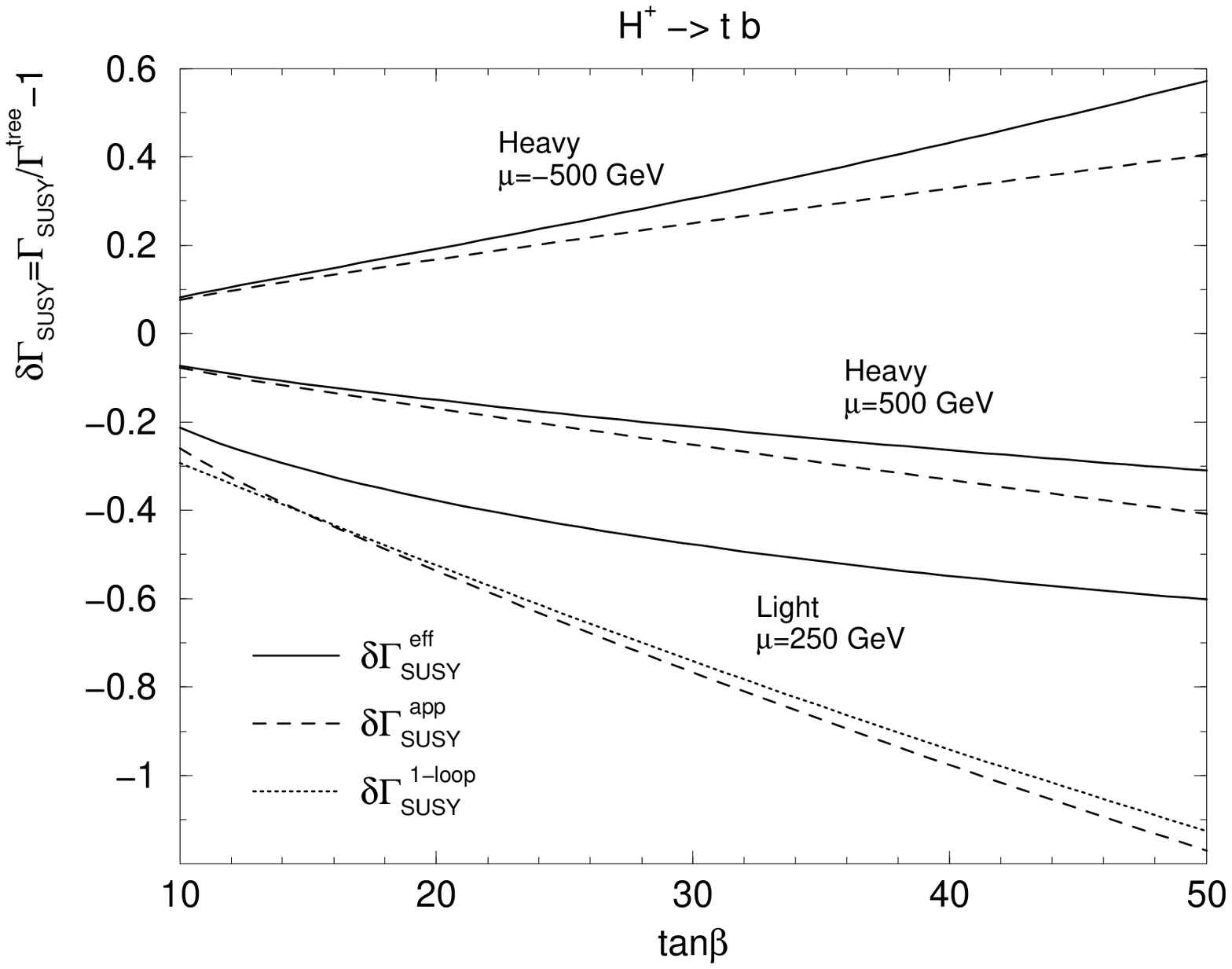,height=7.6cm}
\end{center}
The short-dashed line corresponds to the one-loop result of
\cite{SQCDHtb}. The long-dashed curve is obtained by retaining only
the $\tan \beta$-enhanced contribution from $\delta h_b$. The solid
line supplements the result of \cite{SQCDHtb} with the resummation of 
the large corrections.   

SUSY constraints from $b \to s \gamma$ are highly model dependent.
In the constrained MSSM with $M_{H^+}=200\,$GeV,
$M_{\tilde{t}_2}=250\,$GeV, $M_2=M_{\tilde{t}_1}=M_{\tilde{g}}
=800\,$GeV and $\pm \mu=A_t= 500\,$GeV we\linebreak find \cite{cgnw2}:

\begin{center} 
\psfig{figure=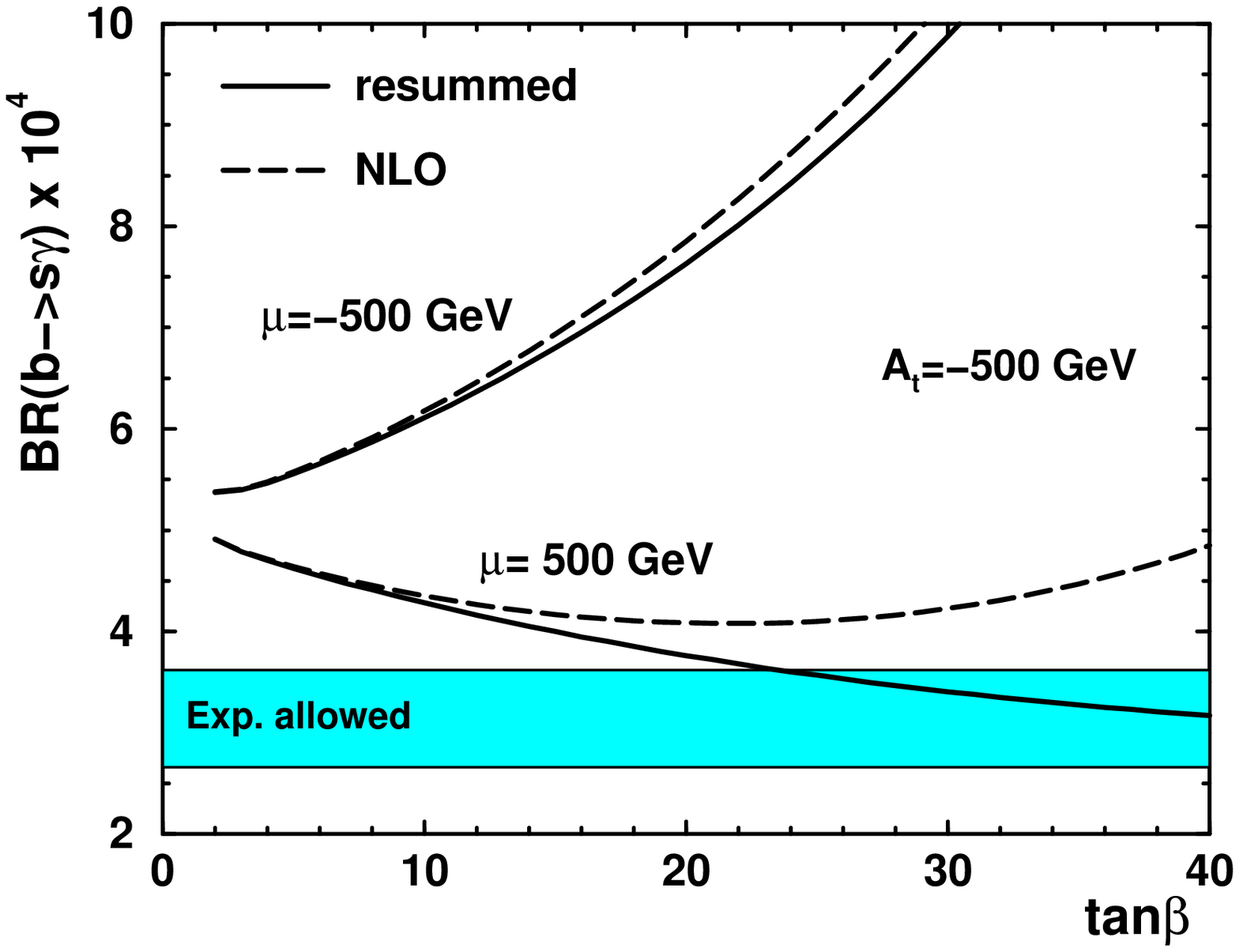,height=7.6cm}
\end{center}
Here the dashed curve is the next-to-leading order result of
\cite{cdgg} and the solid curve represents the resummed prediction for
the decay rate. 
The shaded area is the experimentally \cite{CLEO} allowed range ${\cal BR}(b\to
s\gamma)=(3.14\pm 0.48)\times 10^{-4}$. Using the
resummed result, we can determine the region of the
$(M_{H^+},M_{\chi_2})$-plane excluded by ${\cal BR}(b\to s\gamma)$.
In the following plot we have scanned over
$m_{\tilde{t}_2}<m_{\tilde{t}_1}\leq 1$~TeV,
$m_{\tilde{\chi}^+_2}<m_{\tilde{\chi}^+_1}\leq 1$~TeV and $|A_t|\leq
500$~GeV. The four lines correspond to two values of $\tan \beta$
and of the lighter stop mass as indicated in the plot. The region to the
left of the corresponding line is excluded \cite{cgnw2}:  

\begin{center} 
\psfig{figure=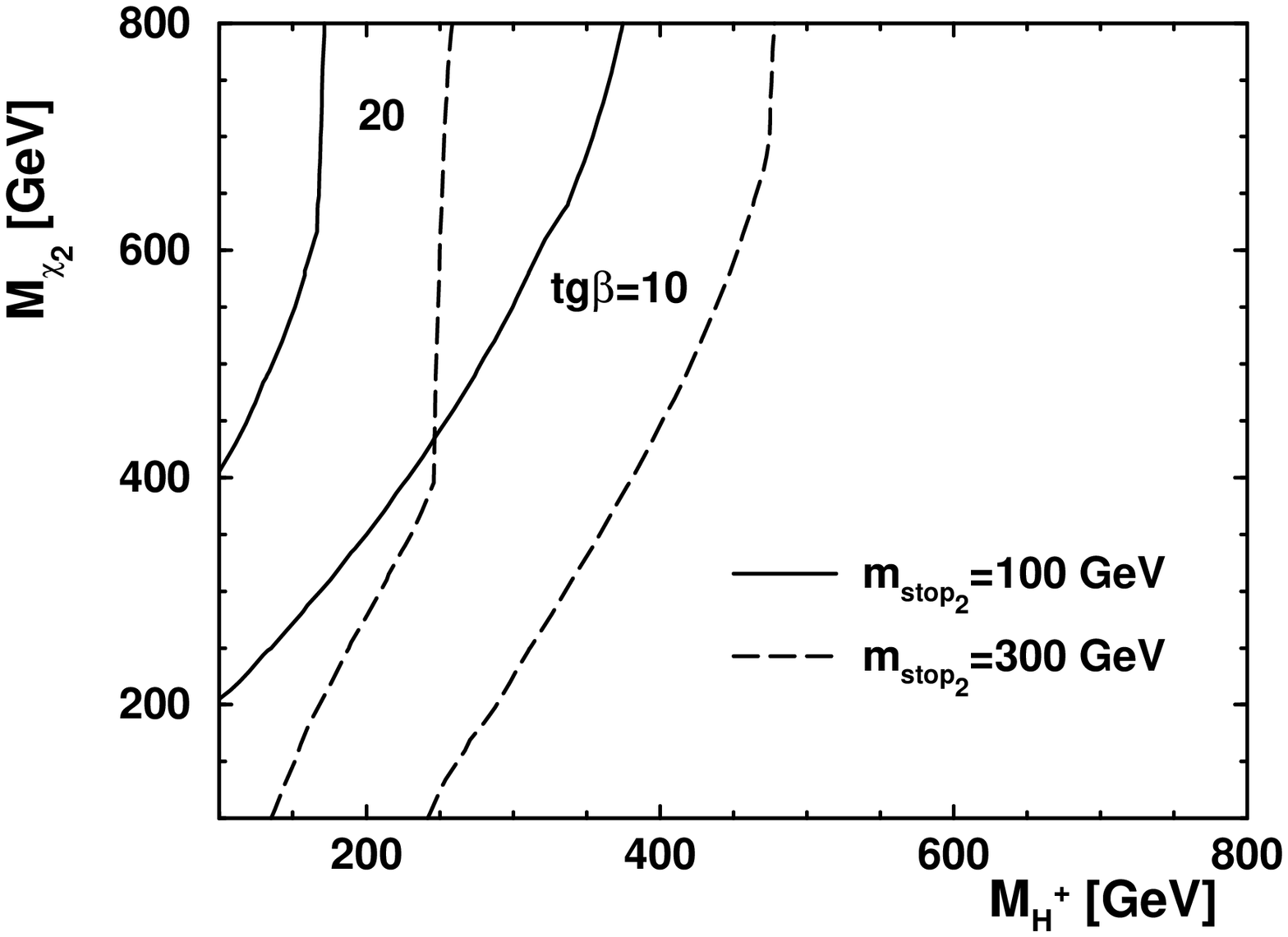,height=7.6cm}
\end{center}
Similar results have been obtained in \cite{dgg}.

\section*{Acknowledgments}
I thank the organizers for creating the stimulating atmosphere of the
conference. Financial support from the European Union is gratefully
acknowledged. I appreciate a very enjoying collaboration with Marcela
Carena, David Garcia and Carlos Wagner on the presented work.


\section*{References}
\begin{thebibliography}{99}
  
\bibitem{lep} ALEPH, DELPHI, L3 and OPAL coll.\ and the LEP Higgs
  Working Group, \emph{Searches for the Neutral Higgs Bosons of the
    MSSM}, L3 Note 2700, July 2001.

\bibitem{gut} 
B.~Ananthanarayan, G.~Lazarides and Q.~Shafi,
Phys.\ Rev.\ {\bf D44} (1991) 1613;
T.~Banks,
Nucl.\ Phys.\ {\bf B303} (1988) 172;
M.~Olechowski and S.~Pokorski,
Phys.\ Lett.\ {\bf B214} (1988) 393;
S.~Dimopoulos, L.J.~Hall and S.~Raby,
Phys.\ Rev.\ Lett.\ {\bf 68} (1992) 1984 and
Phys.\ Rev.\ {\bf D45} (1992) 4192;
G.W.~Anderson, S.~Raby, S.~Dimopoulos and L.J.~Hall,
Phys.\ Rev.\  {\bf D47} (1993) 3702.
\bibitem{gut2}
L.J.~Hall, R.~Rattazzi and U.~Sarid,
Phys.\ Rev.\ {\bf D50} (1994) 7048;
M.~Carena, M.~Olechowski, S.~Pokorski and C.E.M.~Wagner,
Nucl.\ Phys.\ {\bf B426} (1994) 269.

\bibitem{cgnw}
M.~Carena, D.~Garcia, U.~Nierste and C.~E.~Wagner,
Nucl.\ Phys.\ B {\bf 577} (2000) 88
[hep-ph/9912516].

\bibitem{SQCDHtb}
R.A.~Jimenez and J.~Sola,
Phys.\ Lett.\ {\bf B389} (1996) 53;
A.~Bartl, H.~Eberl, K.~Hidaka, T.~Kon, W.~Majerotto and Y.~Yamada,
Phys.\ Lett.\ {\bf B378} (1996) 167.
J.A.~Coarasa, D.~Garcia, J.~Guasch, R.A.~Jimenez and J.~Sola,
Phys.\ Lett.\ {\bf B425} (1998) 329.
J.~Guasch, R.A.~Jimenez and J.~Sola,
Phys.\ Lett.\ {\bf B360} (1995) 47.
J.A.~Coarasa, D.~Garcia, J.~Guasch, R.A.~Jimenez and J.~Sola,
Eur.\ Phys.\ J.\ {\bf C2} (1998) 373.

\bibitem{cdgg} 
M.~Ciuchini, G.~Degrassi, P.~Gambino and G.~F.~Giudice,
Nucl.\ Phys.\  {\bf B534} (1998) 3.

\bibitem{cgnw2}
M.~Carena, D.~Garcia, U.~Nierste and C.~E.~Wagner,
Phys.\ Lett.\ B {\bf 499} (2001) 141.

\bibitem{dgg}
G.~Degrassi, P.~Gambino and G.~F.~Giudice, 
JHEP {\bf 0012} (2000) 009.  
P.~Gambino, talk at this conference.

\bibitem{CLEO}
CLEO Collaboration, CLEO CONF 98-17, ICHEP98 1011.\\
R.~Barate et al. (ALEPH Collaboration), Phys.\ Lett.\ {\bf
      B429} (1998) 169.

\end{thebibliography}
\end{document}